\documentclass[twocolumn]{NobArticle}

\pdfoutput=1

\DefineFNsymbols*{myfoot}{{\textdagger}{\textddagger}{\dagger\dagger}{\ddagger\ddagger}{*}{**}\S\P\|}
\setfnsymbol{myfoot}

\usepackage{amsfonts}%

\usepackage[framemethod=tikz]{mdframed}
\usepackage{fontawesome5}
\definecolor{mygrey}{RGB}{188, 188, 188}
\newmdenv[innerlinewidth=0.5pt,roundcorner=4pt,innerleftmargin=6.25pt,
          innerrightmargin=6.25pt,innertopmargin=4.25pt,innerbottommargin=4.25pt,
          linecolor=mygrey,backgroundcolor=mygrey!25!white,
          skipbelow=0pt,skipabove=0pt,leftmargin=0pt,rightmargin=0pt]{mygreybox}
\usepackage{environ}
\usepackage{varwidth}
\newlength{\mygreyboxWidth}%
\NewEnviron{mygreyboxvar}[1][]{%
    \setlength{\mygreyboxWidth}{\dimexpr%
        +6.25pt%
        +6.25pt%
        +0pt%
        +0pt%
        }%
    \savebox0{%
        \begin{varwidth}{\dimexpr\linewidth-\mygreyboxWidth\relax}%
            \BODY%
        \end{varwidth}%
    }%
    \begin{mdframed}[innerlinewidth=0.5pt,roundcorner=4pt,innerleftmargin=6.25pt,
                     innerrightmargin=6.25pt,innertopmargin=4.25pt,innerbottommargin=4.25pt,
                     leftmargin=0pt,rightmargin=0pt,
                     linecolor=mygrey,backgroundcolor=mygrey!25!white,
                     userdefinedwidth=\dimexpr\wd0+\mygreyboxWidth\relax,
                     skipabove=0pt,
                     #1]
        \usebox0%
    \end{mdframed}%
    \vspace*{-\mdflength{skipbelow}}%
}

\newcommand\blfootnotetext[1]{%
  \begingroup%
  \setlength{\footnotemargin}{0pt}%
  \renewcommand\thefootnote{}
  \footnotetext{#1}%
  \endgroup%
}

\newcommand{\email}[1]{\href{mailto:#1}{\faEnvelope[regular]}}%
\newcommand{\emailr}[1]{\href{mailto:#1}{\texttt{#1}}}%
\newcommand{\orcid}[1]{\href{https://orcid.org/#1}{\faOrcid}}%

\runninghead{Artificial Intelligence Should Genuinely Support Clinical Reasoning and Decision Making To Bridge the Translational Gap}%

\title{Artificial Intelligence Should Genuinely Support Clinical Reasoning and Decision Making To Bridge the Translational Gap}%

\author{
    Kacper Sokol\textsuperscript{1~\orcid{0000-0002-9869-5896}~\email{kacper.sokol@inf.ethz.ch}}, %
    James Fackler\textsuperscript{2,3~\orcid{0000-0003-1377-4366}} 
    and Julia E.\ Vogt\textsuperscript{1~\orcid{0000-0002-6004-7770}} %
}

\date{
    \textsuperscript{\textbf{1}}
    Department of Computer Science, ETH Zurich, Zurich, Switzerland \\ \textsuperscript{\textbf{2}}
    Department of Anesthesiology and Critical Care Medicine, Johns Hopkins University School of Medicine, Baltimore, MD, USA \\ \textsuperscript{\textbf{3}}
    Department of Pediatrics, Johns Hopkins University School of Medicine, Baltimore, MD, USA \\ \textsuperscript{\email{kacper.sokol@inf.ethz.ch}} %
    Corresponding author: \emailr{kacper.sokol@inf.ethz.ch} 
}

\renewcommand{\maketitlehookd}{%
\begin{abstract}

\noindent%
Artificial intelligence promises to revolutionise medicine, yet its impact remains limited because of the pervasive translational gap. %
We posit that the prevailing technology-centric approaches underpin this challenge, rendering such systems fundamentally incompatible with clinical practice, specifically diagnostic reasoning and decision making. %
Instead, %
we propose a novel sociotechnical conceptualisation of data-driven support tools designed to complement doctors' cognitive and epistemic activities. %
Crucially, it prioritises %
real-world impact over %
superhuman performance on inconsequential benchmarks. %

\vspace{5pt}
{\textbf{\normalsize{Keywords:}}\hfill}%
\vspace{2pt}

\noindent%
Artificial Intelligence \textbullet{} Machine Learning \textbullet{} Predictive Modelling \textbullet{} Data-driven Systems \textbullet{} Reasoning Support Systems \textbullet{} Decision Support Systems \textbullet{} Human-centred \textbullet{} Human Decision Making \textbullet{} Augmented Intelligence \textbullet{} Naturalistic Decision Making \textbullet{} Cognitive Forcing \textbullet{} Ecological Validity \textbullet{} Clinical Practice \textbullet{} Medicine \textbullet{} Healthcare \textbullet{} Paediatric Sepsis%
\end{abstract}
}

\begin{document}

\small
\maketitle

\blfootnotetext{%
    \vspace*{-\parskip}%
    \begin{mygreyboxvar}%
\noindent\faIcon*[regular]{file}\hspace{.2cm}%
{\scriptsize\textbf{Published in} %
\emph{Nature npj Digital Medicine} %
(\href{https://doi.org/10.1038/s41746-025-01725-9}{10.1038/s41746-025-01725-9})%
}%
    \end{mygreyboxvar}%
    \vspace*{-\parskip}%
    }%

\section*{Introduction}%

Artificial intelligence (AI) is advancing at a breakneck pace with a promise to overcome numerous real-life challenges across many domains~\citep{sarker2021machine}. %
Of particular relevance is medicine, where %
data-driven tools can %
lead %
to better quality of and access to healthcare -- especially in resource-scarce regions -- %
by %
helping with early detection and prevention of diseases as well as %
delivery of personalised treatments~\citep{chang2020intelligence,bica2021real,johnson2023potential}. %
Specifically, %
AI %
has the potential to %
increase the efficiency of healthcare institutions, %
abate shortages of medical professionals, %
aid with managing the demand for care in view of population ageing and lifestyle diseases, %
alleviate the economic burden of healthcare %
as well as %
reduce the recovery time, mortality and morbidity of devastating medical conditions, saving numerous lives on a global scale. %
However, even the most advanced AI models boasting state-of-the-art or superhuman predictive performance on benchmark tasks have negligible or non-existent benefit -- setting aside the technical progress itself -- if they are never integrated into clinical practice. %
While there have been some success stories in this regard, they remain scarce compared to the sheer number of such systems currently being developed~\citep{gulshan2016development,esteva2017dermatologist,golden2017deep}. %

This phenomenon is a stark manifestation of a \emph{translational barrier} that is ubiquitous in AI for healthcare research~\citep{berg1999patient,wiens2019no,wardi2023bringing,markowetz2024all}. %
While significant focus remains on advancing predictive performance of such models~\citep{wang2007theoretical,komorowski2018artificial,topol2019high,chang2020intelligence,tsirtsis2023finding,wardi2023bringing}, this approach does not appear to offer much progress in terms of AI adoption, except for a very limited range of clinical application domains~\citep{gulshan2016development,esteva2017dermatologist,golden2017deep}. %
Some of the underlying reasons include technical misalignment and incompatibility of such systems with deployment requirements~\citep{berg1999patient}, but frictions at the interface of users and technology as well as societal concerns are more prominent~\citep{wiens2019no}. %
While research into AI fairness, accountability, robustness, interpretability and the like attempts to address these challenges~\citep{qayyum2020secure,mehrabi2021survey,rudin2019stop}, %
progress across these disciplines %
has thus far not managed to unequivocally overcome the translational barrier~\citep{chakravorti2022ai,ghassemi2022medicine,volovici2022steps,tricco2023implemented,sivaraman2023ignore}. %
Since such technological advancements on their own do not seem to deliver the anticipated real-life impact, %
an alternative \emph{sociotechnical} approach focused on seamless integration of AI-based systems into clinical practice and their overall acceptability may be more fruitful~\citep{mueller2019explanation,topol2019high,akata2020research,croskerry2020cognitive,van2021clinical,simkute2022xai,anjum2023conversation,susanto2023effects,wosny2023experience,liao2024ux}. %
After all, even rudimentary data-driven tools that offer modest improvements across a disease lifecycle can have bigger impact than strictly more powerful systems if only the former are adopted by clinicians while the latter remain purely a research feat. %

In this \emph{Perspective} we outline a promising sociotechnical research direction that could help medical AI models to overcome the translational barrier %
by realigning their operation with the needs and expectations of doctors as well as the intricate environments in which these systems operate. %
Achieving these goals requires an \emph{interdisciplinary}, \emph{human-centred} approach that abandons the \emph{autonomous} view of (artificial or human) intelligence and acknowledges its \emph{social} and \emph{relational} nature~\citep{siddarth2021ai,munn2022automation}. %
Instead of striving to replace clinicians with undesired, fallible and potentially harmful data-driven automation, we posit that AI systems ought to seamlessly integrate into and augment -- as opposed to disrupt -- well-established medical workflows as well as real-life reasoning and decision-making processes. %
Consequently, artificial intelligence can assist healthcare professionals in their everyday tasks, complement their abilities, boost their effectiveness and champion clinical best practice~\citep{berg1999patient,sheehan2013informing,winby2018digital,topol2019high,mueller2019explanation,pasmore2019reflections,akata2020research,seeber2020machines,van2021clinical,siddarth2021ai,munn2022automation,shneiderman2022human,wardi2023bringing,miller2023explainable,anjum2023conversation,sivaraman2023ignore,herrmann2023keeping,ferrario2023experts,keenan2023mind,zhou2023artificial}. %

By %
recognising insights from \emph{cognitive sciences} and %
embracing the \emph{systems ecology} of clinical decision making -- that is the complex interconnected network of its various facets -- %
we can design a new generation of AI tools. %
One that %
supports fundamental cognitive (pertaining to conscious intellectual activity) and epistemic (relating to knowledge) functions of doctors -- for example, reasoning under noise and uncertainty -- %
therefore empowers them to make the best judgement given available information. %
Such systems could, among others, %
improve consistency of decisions (e.g., by eliminating decision noise), %
alleviate common reasoning limitations and faults (e.g., arising due to cognitive biases), %
reduce overall (clinical) errors and mistakes (e.g., resulting from a lapse of judgement) %
and generally make the underlying thought processes more principled~\citep{patel1986knowledge,ramoni1992epistemological,kuhn2002diagnostic,croskerry2003cognitive,klein2005five,groopman2007doctors,croskerry2008overconfidence,tetlock2016superforecasting,mueller2019explanation,chang2020intelligence,akata2020research,van2021clinical,simkute2022xai}. %
Our discussion throughout the rest of this \emph{Perspective} is supported by observations and hurdles from (paediatric) \emph{sepsis} %
since this disease offers %
a representative case study of commonplace reasoning and decision-making challenges in medicine; %
nonetheless, the arguments we present generalise to other areas of healthcare as well as different (high stakes) domains. %

Sepsis is a life-threatening condition that arises when human body injures itself in response to an infection~\citep{singer2016third}. %
It is the third leading cause of death (estimated at over ten million a year) and critical illness worldwide, it is the primary cause of mortality from infection and in hospitals, and its survivorship often entails long-term health problems~\citep{fleischmann2018global,morin2022current}, %
not to mention its considerable economic burden~\citep{singer2016third,komorowski2018artificial}. %
However, %
since %
sepsis %
spans a diverse range of incompletely understood processes, it %
remains elusive~\citep{singer2016third}, especially in children for whom it is just as much of a threat as for adults yet it is far less explored~\citep{schlapbach2024international}. %
This %
is particularly concerning given that many observations from the better-understood adult sepsis do not transfer or generalise to the paediatric population with its six clinically and physiologically distinct subgroups~\citep{goldstein2005international,tennant2024scoping}. %

Building upon %
decades of research, %
paediatric %
sepsis %
is currently defined by the rigorous \emph{Phoenix Criteria} established through an international consensus~\citep{schlapbach2024international}. %
Its diagnosis %
is based on suspected or confirmed infection in presence of potentially life-threatening organ dysfunction of the respiratory, cardiovascular, coagulation and/or neurological systems %
determined by %
a \emph{Phoenix Sepsis Score} of two or more. %
Nonetheless, some aspects of this definition remain problematic due to their inherent ambiguity. %
Chief among them is %
predicating sepsis on \emph{suspected} infection, which is %
interpreted as a physician placing an order for a microbiological test; %
but it is well known that such tests are overutilised~\citep{woods2024diagnostic}, likely leading to sepsis overdiagnosis and consequently antibiotic overtreatment~\citep{chiotos2023antibiotic}. %
Presupposing organ dysfunction is also a point of contention as this guideline can be compared to expecting cancer to only be diagnosed after discovering its metastases. %
Further recognition nuances include %
the existence of \emph{culture-negative} sepsis, %
which %
lacks a generally accepted definition but refers to sepsis caused by an infection that is either undetectable by a bacteria culture test %
or simply when this result is \emph{assumed} to be a false negative~\citep{klingenberg2018culture,lee2024comparative}. %

Sepsis is thus %
best described as %
poorly understood. %
Its true incidence is unknown, its best treatment strategy is uncertain and attempts over the past two decades to develop new treatments have been largely unsuccessful %
-- we still lack rigorous clinical criteria, biological markers, imaging features and laboratory tests to identify this disease~\citep{singer2016third,komorowski2018artificial}. %
As it stands, bedside clinicians often struggle to anticipate, identify and treat (paediatric) sepsis given variations in medical guidelines and poor predictive value of many current indicators, creating an urgent need for suitable diagnostic tools that could aid doctors in (less biased and more consistent) decision making and delivery of personalised care~\citep{schlapbach2018defining,tennant2024scoping}. %

\section*{Medical Artificial Intelligence Adoption Challenges}%

Medicine is uniquely positioned to reap the benefits of the recent progress in artificial intelligence given the high impact of even minute improvements in clinical practice~\citep{bica2021real,johnson2023potential}. %
AI tools are of particular importance to the field of paediatrics, where they have been largely underutilised in the recent past~\citep{chang2020intelligence}. %
Building these systems now is especially timely given the increased availability of high-quality, large-scale, real-life data as well as leaps in AI, opening this technology up for many real-world applications~\citep{sarker2021machine}. %

Success stories include automated analysis of medical imaging data -- e.g., detection of diabetic retinopathy~\citep{gulshan2016development}, classification of skin cancer~\citep{esteva2017dermatologist} and detection of lymph node metastases from breast cancer~\citep{golden2017deep} -- enabled by recent advances in deep learning. %
However, such modelling problems are not necessarily representative of medical workflows since they deal with self-contained tasks whose broader context can often be disregarded. %
Additionally, many of these success stories pertain to the visual domain. %
But the accuracy of doctors' diagnostic abilities varies widely between disciplines and tends to be task-specific, with visual specialities -- e.g., dermatology, radiology or anatomic pathology -- exhibiting a far lower error rate (one to two per cent) than many other areas of medicine (around fifteen per cent). %
Such disparities can, among other factors, be attributed to a %
disproportionate signal-to-noise ratio inherent to different specialities~\cite{croskerry2020cognitive}. %

Looking at application of AI-based predictive models and decision-support tools in healthcare through the lens of (paediatric) sepsis offers a more comprehensive perspective. %
In addition to being prototypical, yet unlike other illnesses, this disease is multifaceted and provides enough depth and complexity to elicit real-life desiderata and requirements of such technologies. %
Specifically: %
\begin{itemize}%
    \item sepsis \emph{recognition}
    is problematic due to the ambiguity surrounding relevant definitions (as explained in the previous section); %
        \item its \emph{management}
    is impeded by the lack of well-established and universally accepted tools and techniques for gauging patient risk as well as anticipating the disease progression and its severity; and %
    \item the \emph{treatment} of this illness %
    is inconsistent because of the underlying uncertainty %
    as well as lack of %
    reliable mechanisms %
    to systematically monitor patients' response to therapy (i.e., antibiotics). %
\end{itemize}%
For the paediatric population, its heterogeneity further compounds these issues as they need to be addressed independently for each age group. %

With its diverse open challenges and plentiful avenues for improvement, sepsis offers a perfect case study to stimulate and guide the development of novel medical AI systems~\citep{bica2021real,johnson2023potential,elish2018stakes,tennant2024scoping}. %
While artificial intelligence techniques have been applied to this disease before, real-life impact of such tools is limited. %
Data-driven models %
were used to predict mortality and learn personalised optimal treatment strategies for adults~\citep{komorowski2018artificial,banerjee2021machine,tsirtsis2023finding}; %
the paediatric population, nonetheless, remains largely neglected with only a few studies modelling sepsis onset and mortality~\citep{griffin2001toward,schlapbach2017prediction,joshi2019predicting,tennant2024scoping}. %
More broadly, AI %
was used to predict infection as well as assess susceptibility to antibiotics, quantify exposure to them and optimise their choice~\citep{kanjilal2020decision,moehring2021development,adams2022prospective}. %

Among these, as well as many other, medical AI systems, %
traditional supervised and unsupervised models for classification and regression tasks are the most prominent, e.g., answering questions like ``Will the treatment be continued in five days?'' or ``How many more days will the antibiotics be given?'' %
Such practice, however, %
inadvertently transposes common predictive paradigms %
onto healthcare applications without considering their suitability or adapting them to fit the underlying, well-established clinical reasoning and decision-making workflows and their broader institutional situatedness~\citep{chang2020intelligence}. %
For example, while the evolution of a patient pathway is an inherently continuous process, such a sequence of events is often converted to a classification problem that yields a collection of independent point-in-time predictions about a patient's state in fixed time intervals. %
To illustrate the pitfalls of this modelling approach consider two patients: one whose health is declining and another who is recovering; %
at some point in time their state may be captured by the same data point, therefore they will receive an identical, na\"ive prediction despite one being ready for discharge and the other requiring critical care in the near future. %

AI systems that account for temporality -- thus are able to answer questions such as ``When is the best time to administer antibiotics?'' or ``In how many days will the patient require critical care?'' -- are more appropriate, yet broadly underutilised~\citep{chang2020intelligence,chicco2020survival,liu2023estimating}. %
While the core activity of healthcare professionals is to \emph{manage patients' trajectories} by investigating, monitoring and intervening to palliate and cure medical conditions~\citep{berg1999patient}, AI solutions that support such responsibilities, or simply model patient pathways, are largely missing~\citep{beck2016diagnosis,allam2021analyzing}. %
This is particularly problematic for %
(paediatric) %
sepsis %
since this disease %
is characterised by a change in the patient's condition rather than their absolute health state; %
in case of children, %
this process %
is %
represented by an increase in their %
\emph{Phoenix Sepsis Score}, %
which reflects %
progressive organ dysfunction (as explained in the previous section). %

Consequently, %
despite significant technological advancements, healthcare remains one of the least digitised spheres of life with many open challenges~\citep{berg1999patient,capobianco2019data,spatharou2020transforming,wardi2023bringing}. %
While the number of %
technical solutions proposed in the literature has soared in recent years, such contributions largely focus on developing or adapting general-purpose algorithms to solve narrowly-defined benchmark tasks -- e.g., intensive care unit (ICU) mortality or length of stay -- that are selected primarily based on data availability and evaluation ease. %
These systems are also predominantly optimised for predictive performance with the goal of matching or surpassing capabilities of expert clinicians, boasting impressive results -- often in synthetic or unrealistic experimental settings -- that do not necessarily translate to clinical efficacy or acceptability (among others, due to their inherent misalignment with medical practice)~\citep{wang2007theoretical,topol2019high,wardi2023bringing}. %

While artificial intelligence -- as a technology -- is agnostic of its application, %
it is fundamentally not a one-size-fits-all solution; %
applying \emph{generic} data-driven algorithms to medical challenges simply because relevant data are available is thus unlikely to deliver useful tools~\citep{berg1999patient}. %
Healthcare requires bespoke AI models tailored to each unique clinical application, especially since incorporating domain-specific knowledge into them tends to enhance their performance and improve their acceptability~\cite{wiens2019no}. %
The lack of such considerations %
leads to %
a mismatch between development/validation and deployment contexts or desiderata, preventing AI tools from being integrated into clinical workflows or simply making them unusable in practice~\citep{berg1999patient,tennant2024scoping}. %
In case of technical requirements, for example, %
it is common to presuppose access to biomarkers from medical test results at the time of acquiring a sample as opposed to receiving the corresponding lab report, thus allowing AI to rely on information from the future, which disqualifies it from real-time operation~\citep{komorowski2018artificial}. %

When such ill-conceived AI systems -- whose functioning is at odds with practical constraints -- are deployed, they are often %
ignored or dismissed by clinicians because of general frustration as well as mistrust, apprehension or aversion towards their outputs~\citep{tricco2023implemented,sivaraman2023ignore}. %
In part, these attitudes can be attributed to AI %
being overly time-consuming to use, %
entailing high cognitive burden, %
failing to deliver the necessary information, %
disrupting or not integrating well into existing decision-making workflows %
and the like~\citep{bansal2021does,simkute2022xai,anjum2023conversation}. %
The lacklustre adoption of such tools is further compounded by pervasive reproducibility issues, %
history of unsafe systems being deployed prematurely, %
prevalence of false automation promises as well as %
scarce data that are inherently private, difficult to collect, store, access or share, and often riddled with numerous biases~\citep{gibney2022could,ghassemi2022medicine,sohn2023reproducibility}. %
Ethical concerns and negative societal consequences of deploying data-driven predictive models in healthcare -- where they may cause direct harm on a large scale -- also stymie their integration into clinical practice. %
While the algorithmic nature of such techniques streamlines decision-making processes and arguably makes them more objective and equitable by replacing biased and fallible humans~\citep{kahneman2009conditions}, these anticipated benefits have time and again not come to fruition or been overshadowed by unforeseen adverse societal impact, disproportionately affecting minorities and people of colour~\citep{o2016weapons,angwin2016machine,dastin2018amazon,carney2019robo,wiens2019no,volovici2022steps,chakravorti2022ai,geiger2023suspicion}. %

From historical biases and discrimination captured in data and perpetuated by automated decisions, through entire populations being underrepresented in training samples leading to predictive performance disparities and unfairness, to modelling reliant on spurious data patterns, building AI models without unintended consequences is a formidable challenge~\citep{o2016weapons,angwin2016machine,barocas2017fairness,buolamwini2018gender}. %
Fairness considerations are particularly pertinent in medical applications where protected attributes such as income, gender or ethnicity may be good (proxy) clinical predictors. %
Healthcare is a high stakes domain that requires thoroughly validated, fair, privacy-preserving, interpretable, reliable, robust and accountable AI systems; %
they additionally must %
satisfy the necessary regulatory requirements
as well as be
compatible with clinical workflows and
suitable for real-time operation~\citep{qayyum2020secure,mehrabi2021survey,rudin2019stop,chang2020intelligence,kahneman2021noise}. %
Deploying and maintaining data-driven models to keep them functional, usable and relevant can also be a hurdle given resource constraints like computational infrastructure requirements and running costs. %
All of these aspects contribute to the aforementioned \emph{translational barrier} -- a chasm between technical solutions and clinical applications -- which while pervasive is sporadically documented because AI is rarely tested (prospectively) in real-life clinical settings~\citep{berg1999patient,komorowski2018artificial,wiens2019no,kanjilal2020decision,moehring2021development,adams2022prospective,tsirtsis2023finding,wardi2023bringing,markowetz2024all}. %

\section*{Systems Ecology and Artificial Intelligence}%

The factors outlined in the previous section often lead to data-driven tools that are impractical and lack real-world usability, which hampers their adoption~\citep{simkute2022xai,anjum2023conversation,tennant2024scoping}. %
This phenomenon is further compounded by %
the challenge of defining the epistemic task that AI should solve; %
consequently, %
predictive models are primarily trained to mimic human decisions since this strategy offers a tractable proxy objective %
that allows for direct optimisation of predictive performance as well as evaluation and benchmarking against human experts. %
However, such a reductionist approach that (over)simplifies the role of technology in its designated real-world environment often leads to pervasive lack of \emph{ecological validity}~\cite{croskerry2020cognitive}. %
The umbrella term of data-driven \emph{decision support} tools used to describe these systems is thus a misnomer since they are rarely designed to genuinely \emph{support} decision making and instead present users with ready-made conclusions that compete with and curtail their own judgement~\citep{van2021clinical}. %
The \emph{support} aspect, if any, is mostly confined to justifying predictions with algorithmic explanations and embedding this process within a \emph{human-in-the-loop} interaction protocol that seemingly blends human and machine decisions~\citep{susanto2023effects,anjum2023conversation,miller2023explainable}. %

This %
fundamental misalignment between the operationalisation of AI and human decision making leads to undesired automation that %
biases perception, %
impedes cognition, %
limits independent reasoning, %
inhibits natural exploration and %
hinders sense making~\citep{wosny2023experience,anjum2023conversation,byrne2023good}. %
These factors tend to contribute to unwarranted reliance on AI and automation bias, %
but more notably they erode the value of expertise, %
disrupt well-established workflows as well as disempower, disenfranchise and displace people instead of supporting them and augmenting their abilities. %
In the worst case, for predictive models integrated into clinical practice, eliciting such behavioural patterns %
can reinforce cognitive blunders of doctors, thus undermine the care they provide~\cite{croskerry2020cognitive,jin2023rethinking}. %
This is especially worrying in the context of the \emph{medical diagnostic error} being estimated as one of the most consequential, and often a leading cause of (preventable) death~\cite{croskerry2003cognitive,croskerry2020cognitive}. %
Notably, the majority of such errors are not rooted in insufficient medical knowledge or inadequate expertise, but rather in \emph{structural causes} that result in deficiencies of medical judgment, %
with misdiagnosis rate reaching twenty-three per cent in everyday practice for healthcare domains that necessarily rely on high levels of subjectivity, where interobserver variation in diagnosis is to be expected~\citep{kuhn2002diagnostic,graber2012cognitive,battefeld2022formalizing}. %

The aforementioned structural factors include, among others, time pressure, uncertainty and various cognitive biases, %
all of which lead to problematic synthesis of diagnostic information. %
The two most prevalent causes are %
\emph{anchoring} -- i.e., steadfastly sticking to an initial impression, thus possibly ignoring subsequent evidence that may be contradictory or disproving -- and \emph{premature closure} -- i.e., jumping to a conclusion without considering all the available or necessary evidence. %
These biases seem to %
affect doctors regardless of their experience, but they appear more common in experts~\citep{kuhn2002diagnostic}. %
While medical reasoning and diagnostic error is amply documented in retrospect, %
detecting and preventing it prospectively is inherently difficult, and largely underexplored in the literature, because it does not manifest as openly as practical mistakes. %
As a consequence, any data that capture such aspects of clinical work may implicitly encode outcomes of inconsistent or incorrect diagnostic reasoning, with the ensuing AI models inadvertently perpetuating these errors. %

In view of these observations, overcoming the translational barrier is likely to require a fundamental change in the design and implementation of artificial intelligence systems. %
Specifically, their creation %
ought to be motivated by concrete needs, requirements and challenges faced by human decision makers, helping people to overcome their limitations while also eliciting their strengths~\citep{berg1999patient,anjum2023conversation}. %
Additionally, the integration of these tools into decision-making workflows should not only be informed by viewing humans as independent agents, but also by recognising their role and placement in the broader context of the processes and environmental constraints in which they operate~\cite{topol2019high,mueller2019explanation,ferrario2023experts,akata2020research,keenan2023mind,tennant2024scoping}. %
Such AI systems have the potential to %
augment specific cognitive tasks, %
boost comprehension, %
promote active exploration, %
stimulate creative problem solving and %
facilitate (prospective) critical reasoning, %
thus truly aid and support evidence-driven decision making instead of attempting to ``solve'' it algorithmically through disruptive automation. %
To this end, artificial intelligence could, for example, %
fill the gaps in people's \emph{knowledge}, %
allow individuals to challenge automated decisions and then help them to consider and compare alternatives via AI-assisted prospective \emph{mental simulation}, or %
aid people in progressively \emph{updating their beliefs} to arrive at sound conclusions and decisions~\citep{mueller2019explanation}. %

The \emph{technological} translational barrier described in the previous section should therefore also be viewed as a \emph{sociotechnical gap} -- i.e., the differentiation of what must be supported socially and what can be supported technically~\citep{ackerman2003sociotechnicalgap,ehsan2023charting,keenan2023mind} -- overcoming which requires an interdisciplinary approach that draws insights from social and cognitive sciences~\citep{berg1999patient,sheehan2013informing,patel2022clinical,wardi2023bringing}. %
The aforementioned (counterproductive) drive to match or exceed human-level performance in selected (often narrowly- or ill-defined) tasks with the aim of fully automating and replacing humans is thus a manifestation of AI systems being commonly misconstrued as ``autonomous rather than social and relational''~\cite{siddarth2021ai}. %
This %
paradigm -- sometimes referred to as \emph{the race to the bottom} given its intention to remove agency and decision-making power from (individual) humans, shifting the authority to AI and its developers -- is nonetheless challenged ever more frequently~\citep{mueller2019explanation,munn2022automation}. %
Replacing humans with artificial intelligence may not yield the anticipated level of automation but instead shift people from (meaningful and engaging) decisive positions to (frustrating and dreadful) supervisory roles, by and large depriving them of any autonomy and (collective) bargaining power (by making them appear redundant)~\citep{brynjolfsson2022turing,munn2022automation,miller2023explainable}. %
Such a reconfiguration is particularly ironic for high stakes domains where lifting the perceived limitations and failures of humans with AI often requires those same flawed humans to monitor, interpret, vet and correct computers' output~\citep{bainbridge1983ironies}. %

In addition to possible bias, discrimination, unfairness and displacement, %
full automation also raises ethical concerns given unclear attribution of responsibility when an algorithmic decision causes unintended harm. %
In contrast, the responsibility remains with humans when instead of replacing them, AI augments and aids their decision making by providing them with supporting information to be utilised within well-established frameworks like \emph{evidence-based medicine}~\citep{van2021clinical}. %
Similarly, while ignoring or overriding decisions of AI whose performance is supposedly superior to that of human experts could be considered malpractice~\citep{hacker2020explainable}, %
such claims are often technically dubious (as argued earlier) and arise primarily from the autonomous view of intelligence~\citep{ferrario2023experts}. %
This is not to say that (full automation based on) predictive systems %
should be discarded altogether, %
but rather that %
the adoption of data-driven tools ought to be based on sound justification and robust defence of the process and means via which scientifically grounded and rationally defensible decisions are reached~\citep{van2015epistemological}. %

Consequently, integration of AI cannot be premised on \emph{trust}, e.g., established over prolonged interaction episodes~\citep{tomsett2020rapid}, %
as this concept is ill-suited for technology~\citep{ryan2020ai}; %
instead, artificial intelligence should be judged in terms of \emph{reliability} and \emph{robustness}. %
After all, ``if the outcome of a traditional machine becomes unpredictable, we do not think that it is creative or original -- we think that it is broken''~\citep{esposito2022artificial}, and AI should be treated no different. %
These observations reinforce the relevance of \emph{ante-hoc interpretable} artificial intelligence whose inherent soundness and human intelligibility are guaranteed by means of constraining the underlying model form, e.g., to account for application-specific requirements, making it suitable for high stakes domains like healthcare~\cite{rudin2019stop,rudin2022interpretable}. %
This AI transparency paradigm is distinct from more prevalent \emph{post-hoc explainability}, %
which simplifies opaque predictive systems to make them human-understandable %
by approximating their operation. %
The insights output by such methods, however, %
are not guaranteed to reflect the true behaviour of the underlying models; they may thus be misleading, hence unacceptable in some (safety-critical) applications~\citep{rudin2019stop}. %
Out of the two, it is ante-hoc interpretability that %
delivers a solid foundation for building human-centred predictive systems %
that are \emph{acceptable} -- as a result of appropriate social structures -- \emph{reliable} -- because of sound technical practices -- and \emph{safe} -- due to open management strategies~\citep{shneiderman2020human}. %

In addition to recognising people as \emph{independent agents} interacting with or being replaced by AI, they should also be viewed holistically as members of various societal and organisational structures that connect diverse stakeholders and facilitate their seamless communication and collaboration~\cite{keenan2023mind}. %
To avoid disrupting the fragile systems ecology, (interdisciplinary) AI development teams need to identify the best way of integrating this technology into such complex real-life settings; %
relevant considerations include %
ensuring its compatibility with environmental constraints, established communication protocols as well as institutional interdependencies, workflows, processes, objectives, desiderata, requirements, regulations, (industry) standards and best practice~\cite{berg1999patient,sheehan2013informing,winby2018digital,topol2019high,pasmore2019reflections,seeber2020machines,tennant2024scoping}. %
Within this landscape, one must account not only for the relation between AI and individuals, but also their groups and the overarching organisations, striving to understand the roles, responsibilities and needs of each unit: %
how it works, interacts with other units, processes or exchanges information and makes decisions~\citep{herrmann2023keeping}. %
Many such interconnected systems are set up to provide operational frameworks that streamline task execution by division of labour and responsibility. %
This arrangement %
allows each unit to treat parts of the process as black boxes (which may only appear so while in principle being comprehensible with suitable expertise~\citep{sokol2023reasonable}) that are robust and reliable given the existence of organisational mechanisms to ensure their proper functioning, manage risk and absorb contingencies~\citep{keenan2023mind}. %

By considering the broader societal context %
when building AI tools, %
an opportunity arises %
to assist, enhance, support and enable humans to flourish and excel at their work %
-- a strategy that appears %
more promising than simply attempting to replace them. %
This vision %
can be realised not only with full automation of selected tasks, but crucially through human--machine symbiosis, collaboration, co-creation or hybrid intelligence~\citep{berg1999patient,topol2019high,akata2020research,siddarth2021ai,shneiderman2022human,wardi2023bringing,miller2023explainable}. %
In view of the systems ecology, implementing AI within the \emph{distributed cognition} paradigm -- where distinct responsibilities are allocated to specialised, algorithmic or human, agents -- can be highly beneficial, allowing to support people in their various cognitive and epistemic activities, e.g., comprehension, sense making, reasoning, problem solving, decision making and task execution~\citep{parasuraman2000model,mueller2019explanation,ferrario2023experts}. %

As an example, consider doctors' reliance on and widespread integration into clinical workflows of laboratory test results or outputs of advanced medical devices like magnetic resonance imaging (MRI). %
While in principle these instruments are not black boxes, they can be safely used as such, %
without in-depth understanding of their inner workings or the underlying chemistry, physics or signal processing principles. %
This is possible because the responsibility for correct and reliable functioning of these tools has been shifted to appropriate entities: industries (e.g., tasked with device construction), government bodies (e.g., overlooking its certification processes) and professional staff (e.g., entrusted with its calibration and operation)~\cite{keenan2023mind}. %
Consequently, the existence of these organisational structures streamlines %
the day-to-day work of doctors. %

In medicine, but also elsewhere, AI is perceived no different to such tools~\citep{akata2020research}. %
Its reception is %
more favourable when it is provided as a digital \emph{partner} that complements, augments, amplifies and supports people's abilities and decision making -- e.g., by adding more evidence, compensating for human weaknesses, preventing common biases and overcoming human limitations -- rather than replacing human intelligence or reducing the role of people to accepting/rejecting algorithmic recommendations~\citep{berg1999patient,sivaraman2023ignore}. %

\section*{Human Decision Making and Artificial Intelligence}%

To overcome the challenges outlined in the previous section, it is first necessary to understand %
how experts make decisions in highly structured environments %
and place this process in the context of human--AI dynamics. %
Under the assumption that replacing people or competing with them is counterproductive at best and harmful at worst~\citep{siddarth2021ai,munn2022automation}, we can identify different modes of artificial intelligence operating vis-\`a-vis humans~\citep{chang2020intelligence}. %
While many fine-grained taxonomies exist~\citep{sheridan1978human}, distinguishing the following three levels of AI integration suffices for our purposes. %

\begin{description}[font=\normalfont\bfseries]

\item [autonomy] %
Tasks that do not require direct human input %
can be automated, with the resulting artefacts integrated into higher-level workflows as \emph{additional sources of information} or treated as \emph{prescriptive decisions} only to be monitored by people. %

\item [assistance (human-in-the-loop)] %
Tasks that require human input can benefit from \emph{descriptive} (summarising information and extracting insights of interest) or \emph{predictive} (forecasting quantities) \emph{modelling}, %
with the resulting artefacts supporting %
problem understanding and decision making. %
To this end, the %
insights produced by AI are %
incorporated into the corresponding %
workflows as \emph{clues}, \emph{diagnoses} or \emph{recommendations} to be reviewed and accepted or rejected by people. %

\item [augmentation (machine-in-the-loop)] %
Tasks whose full or partial automation is undesired or infeasible, %
thus confining them to the purview of humans, %
can be streamlined by supporting the higher-level cognitive and epistemic functions of people responsible for their completion. %
To this end, descriptive and predictive modelling can be integrated into a collaborative co-creation process in which AI systems complement humans' abilities and help them to overcome personal shortcomings and limitations, e.g., by presenting people with a range of possible choices accompanied by their respective positive and negative consequences. %
\end{description}

Notably, the most suitable automation paradigm should be selected independently for each separate (cognitive and epistemic) activity such as \emph{data acquisition}, \emph{information analysis} as well as \emph{decision} or \emph{action selection} and \emph{implementation thereof}~\citep{parasuraman2000model,phillips2013intelligent}. %
The specific form in which AI can be safely and responsibly operated ought to be determined based on the \emph{automation readiness level} of a particular task -- a concept that can already be found in data science (\emph{data readiness levels}~\citep{lawrence2017data}), autonomous driving (\emph{levels of automation}~\citep{nhtsa2022levels}) and digital healthcare (\emph{levels of maturity} prescribed by the \emph{analytics adoption model}~\citep{himss2021adoption}). %
Implementing data-driven tools in practice, however, remains challenging. %
This is because %
we generally lack the corresponding (technical) frameworks, guidelines and protocols that would formalise and lead the development and deployment of these systems. %
While true for many domains, including medicine~\citep{volovici2022steps}, %
a notable exception is data mining with its CRISP-DM process~\citep{martinez2019crisp}. %

This perspective on AI integration is compatible with cognitive and behavioural psychology research, which %
offers two complementary viewpoints on human decision making: \emph{Naturalistic Decision Making} studies the success of expert judgement, whereas \emph{Heuristics and Biases} (commonly known as the dual-process theory or System~1 and System~2 thinking~\citep{kahneman2011thinking}) deals with faults in basic reasoning~\citep{kahneman2009conditions}. %
Findings from these disciplines suggest that people's ability to become good decision makers %
as well as the viable level of automation of a given task %
depend largely on the properties of the domain in which such processes take place. %
Settings referred to as %
\emph{wicked environments} either offer misleading signals or lack reliable cues, regularities and feedback for people to observe, learn from and develop correct and complete intuitions~\citep{hogarth2001educating,kahneman2009conditions}. %
But where humans flourish, their actions can be studied (through \emph{Naturalistic Decision Making} frameworks) to possibly identify the source of their expertise and codify this knowledge in textbooks or predictive algorithms. %
In domains %
where people fail, AI may be able to learn and distil patterns that humans cannot, and use them to make decisions or present them to people in a digestible format~\citep{crandall1991guide,crandall1993critical,hogarth2001educating,phillips2013intelligent}. %

Through such an approach we can recognise environments that offer \emph{sufficient regularities} to be amenable to full or partial automation. %
This operationalisation of AI is %
best suited for procedural and repetitive tasks that both machines and humans can complete (on their own) as well as challenges that people struggle with but (data-driven) algorithms can streamline or outright solve. %
In the former case, the benefit comes from reducing (costly) human errors that arise due to a lapse of judgement, yielding improved (predictive) performance attributed to better decision consistency and efficiency. %
In the latter scenario, automation makes up for human cognitive deficiencies in tasks that are inherently incompatible with our reasoning capabilities or simply too complex for our minds. %

Crucially, both of these AI deployment scenarios presuppose that the \emph{stable-world principle} -- sometimes referred to as the \emph{closed-world principle} -- holds and that the selected task is \emph{structured}. %
The former premise %
implies that %
the decision-making environment is inherently predictable (given a set of observations) and that it does not evolve unexpectedly (over time), e.g., resulting in a data shift~\citep{katsikopoulos2021classification,ferrario2023experts,gigerenzer2023psychological}. %
Regarding the latter tenet, we can generally %
distinguish three decision categories~\citep{phillips2013intelligent}: %
\begin{itemize}
    \item 
\emph{structured}
tasks -- referred to before -- come with a well-defined problem that %
has a \emph{single} optimal, possibly analytical, solution, %
which in principle can be found; %
    \item 
    for
\emph{unstructured} %
challenges, %
\emph{no} universally %
accepted
solution exists given that it depends on individual preferences; and %
    \item 
\emph{semi-structured} %
problems have %
\emph{multiple} viable solutions -- %
each with its own pros and cons --
determined according to %
some predefined criteria %
and ranking them requires analytical methods, which may include AI, as well as the decision maker's input. %
\end{itemize}

However, evidence-based medical diagnostic reasoning %
often necessarily relies
on incomplete and uncertain information, with some areas of healthcare requiring a high level of subjective judgement and providing outcomes that may not always be fully predictable~\citep{dowding2009evidence,battefeld2022formalizing}. %
This is exactly why evidence-based medicine refrains from prescriptive rules and instead offers best practice guidelines; %
a decision-making environment as complex as healthcare cannot be easily distilled into rigid procedures that anticipate and encompass all the unique circumstances of individual patients~\citep{van2015epistemological}. %
Consequently, semi-structured clinical tasks may not be amenable to (end-to-end) AI modelling %
as one ``optimal'' decision or solution that these systems tend to deliver is unlikely to exist in this context. %
Given the assertive nature of such algorithmic recommendations, they can also stymie %
human-driven exploration and discovery of alternatives that may prove better in the long term, %
possibly curtailing the progress of medicine~\citep{phillips2013intelligent}. %
More generally, since predictive models usually optimise for past outcomes, %
their adoption inadvertently risks hampering scientific serendipity as well as impeding development of new and advancement of existing knowledge~\citep{jin2023rethinking}. %

In a stable-world setting, large data quantities and advanced learning algorithms tend to offer unparalleled performance for structured and semi-structured problems (e.g., in the game of chess or go). %
For open-world tasks, however, simple and inherently transparent models or high-level decision heuristics (both of which can be seen as forms of ante-hoc interpretable AI) can perform on a par with or better than complex data-driven systems (e.g., for predicting heart attack risk)~\citep{gigerenzer2023psychological}. %
Humans are known to rely on such straightforward heuristics and biases in their everyday decision making. %
When studied in a laboratory setting, these mental processes give rise to seemingly suboptimal, irrational or faulty reasoning as reported by the \emph{Heuristics and Biases} community. %
Yet when viewed as evolutionary adaptation necessary to deal with complex and \emph{unstable} (\emph{open-world}) decision-making environments fraught not only with risk but also high degree of unpredictability and uncertainty, these reasoning patterns tend to manifest \emph{ecological rationality} rather than universal defects of cognition~\citep{gigerenzer2023psychological}. %
While such aspects of human decision making are largely overlooked by AI research, they can inspire the design of predictive models -- suitable for stable- and unstable-world problems -- that offer better utility and acceptability than what is currently available. %
Among others, this new class of AI systems could help people to %
boost their comprehension and reasoning abilities in complex environments, increase decision consistency, overcome detrimental cognitive biases, reduce different types of errors and improve overall decision hygiene. %

Instances of both reliable and wicked (e.g., due to their unstable and open-world nature) environments can be readily found in medicine. %
For example, %
nurses in neonatal ICUs were shown to correctly identify infants developing life-threatening infections (leading to paediatric sepsis) without knowing blood test results, yet they were unable to describe or explain their reasoning~\citep{crandall1993critical}. %
The \emph{Naturalistic Decision Making} framework has been applied to study individual incidents and uncover the cues, patterns and observations that the nurses relied on, which led to the discovery of novel insights -- including infection indicators opposite to those relevant for adults -- validated across different hospitals and formalised into an instructional programme to help medical staff spot early signs of neonatal sepsis~\citep{crandall1991guide}. %
On the other hand, %
the prevalent uncertainty surrounding various aspects of this disease and its treatment strategy (discussed earlier) is a clear manifestation of the underlying environment being inherently wicked. %
These circumstances contribute to doctors taking different actions in similar situations spread over time~\citep{al2021context} %
-- a phenomenon that in cognitive psychology is called decision \emph{noise}, defined %
as ``undesirable variability in judgments of the same problem''~\citep{kahneman2021noise}. %

In sepsis, %
early intervention %
should significantly lower the risk of mortality and morbidity, %
especially so for the paediatric population in which infections can be extremely fulminant~\citep{stocker2023less}. %
This belief, for example, motivates research efforts %
to swiftly identify \emph{serious bacterial infections} so that clinicians can intervene \emph{before} patients develop severe organ dysfunction~\citep{martin2022machine}. %
Rapid medical response is thus considered best practice, %
prompting doctors to administer the most effective treatment -- antibiotics -- %
both when the underlying infection is \emph{confirmed} (i.e., clinically proven) or simply \emph{suspected} (i.e., for individuals at risk but not necessarily septic or when \emph{culture-negative sepsis} is surmised)~\citep{cabral2015s,schlapbach2018defining,fontela2022clinical}. %
Such an approach, however, %
leads to many patients -- more than ninety-eight per cent of neonates by some estimates -- receiving an unnecessary or longer than required treatment, e.g., when the underlying illness is self-limited or infection is disproven~\citep{stocker2023less}. %
Consequently, sepsis faces yet another challenge: %
antibiotic overtreatment and its plentiful long-term dangers~\citep{llor2014antimicrobial}. %

Trading off outcomes and time by preferring immediate results -- e.g., the perceived safety offered by antibiotics -- in favour of potential future benefits -- e.g., preventing antimicrobial resistance -- is captured by the \emph{time preference theory}~\citep{frederick2002time,mahboub2014stated}. %
This ``bias towards short-term rewards'' together with confirmation bias, entrenchment, in-group favouritism, limited attention span and restricted short-term (working) memory are just some cognitive biases and heuristics that may contribute to humans inadvertently making suboptimal decisions~\citep{kuhn2002diagnostic,klein2005five,groopman2007doctors,akata2020research}. %
Supporting doctors' cognitive and epistemic functions with AI can thus prove more beneficial than building predictive models optimised to (na\"ively) mimic and improve upon their actions~\citep{tennant2024scoping}. %

\section*{From Benchmark to Bedside}%

Investigating the complex and challenging environment posed by sepsis -- in which clinicians struggle to consistently make optimal decisions -- inspires a different use of artificial intelligence. %
One where its operation better aligns with %
the underlying clinical processes, human decision-making workflows and the broader institutional situatedness thereof; %
and one where its functioning %
respects the abilities of doctors and caters to the needs of their cognitive and epistemic activities. %
Adhering to these principles has the potential to deliver a technology that is readily accepted and adopted by clinicians, especially if it %
embraces, and does not disrupt, the overarching systems ecology, hence %
seamlessly blends into existing structures instead of being provided as a standalone tool~\citep{cook2008human,tennant2024scoping}. %
Since artificial and human intelligence each exhibits distinct capabilities, which tend to complement one another, the former can augment the latter (instead of replacing it) in the form of hybrid intelligence~\citep{akata2020research,van2021clinical}. %
This integration may span different stages of reasoning and decision making that arise throughout patient care, e.g., perception, comprehension, cognition and operation, addressing their respective unique desiderata~\citep{parasuraman2000model,croskerry2003cognitive,chang2020intelligence,corti2024moving}. %
In particular, we can draw inspiration from insights, tools and techniques in cognitive sciences that help human experts make good and reliable decisions in highly structured environments such as clinical practice~\citep{groopman2007doctors,graber2012cognitive}. %

On a high level, these approaches strive to increase the knowledge and expertise of doctors, offer them situational help as well as improve their critical thinking and reasoning processes~\citep{graber2012cognitive}. %
Crucially, these techniques can be naturally embodied and enhanced by AI tools with the aim of providing advanced cognitive support, preventing common biases, alleviating decision-making challenges, overcoming human cognitive limitations and delivering contextual data-driven insights. %
Such intelligent systems promise to make %
up for the shortcomings of these processes that commonly arise due to various environmental factors %
and improve the overall decision hygiene %
on multiple levels~\citep{croskerry2003cognitive,akata2020research,simkute2022xai}. %
Rudimentary research in this direction has explored %
the influence of selected AI explanation types on %
human cognitive biases, constraints and reasoning faults, demonstrating that while some can be mitigated, others may inadvertently be exacerbated~\citep{wang2019designing,buccinca2021trust,bertrand2022cognitive}. %
However, this line of work by and large overlooks the broader systems ecology and remains confined to explainability of predictive models that mimic human decision making~\citep{sokol2025nexai}.%

But in this context, AI integration possibilities are much broader, %
with explainability research poised to offer a viable implementation framework~\citep{sokol2025nexai}. %
Among others, artificial intelligence tools could~\citep{patel1986knowledge,ramoni1992epistemological,croskerry2003cognitive,klein2005five,croskerry2008overconfidence,tetlock2016superforecasting,mueller2019explanation,van2021clinical}: %
\begin{itemize}
    \item 
distil high-level human-comprehensible concepts from data, aid in pattern recognition as well as generate and test hypotheses; %
    \item 
facilitate ideation and evaluation of ideas as well as forward projections and prospective reasoning by supporting mental simulation; %
    \item 
prompt reasoning by analogies and counterexamples as well as identify incongruent, ambiguous or atypical manifestations of modelled phenomena (e.g., when evidence captured by data does not align); %
    \item 
clearly indicate the context of each output by grounding it in domain knowledge, emphasising relevant clinical guidelines and offering pieces of knowledge that the target audience may be lacking (e.g., information about drug interactions); and %
    \item 
stimulate metacognition through self-monitoring, self-critique as well as self-policing in the long term. %
\end{itemize}

One such specific approach is %
\emph{cognitive forcing}, which encompasses an array of (intervention) techniques intended to disrupt heuristic reasoning and trigger analytical thinking, thus %
prompt people to account for overlooked or disconfirming evidence, competing hypotheses and opposing ideas. %
These methods reduce overconfidence, decrease reliance on hunches and intuitions as well as improve reasoning quality and decision reliability~\citep{croskerry2003cognitive,croskerry2008overconfidence}. %
Another relevant strategy is to %
facilitate and encourage continuous, as opposed to one-off, decision making, where people explicitly account for the incidence of a given phenomenon (i.e., its base rate) and progressively \emph{update their beliefs} by considering standalone insights along with their relevance/salience~\citep{tetlock2016superforecasting}. %
Moreover, humans tend %
to perform better over time when they are provided with (immediate) \emph{feedback} or are prompted to perform post-factum evaluation of situations that they have encountered and decisions that they have taken. %
This approach
together with the other aforementioned decision-making strategies %
facilitate and stimulate long-term learning and expertise development. %

These mechanisms can, for example, be delivered as part of an AI toolkit that %
supports clinicians in \emph{mental simulation}. %
Specifically, this type of technology could %
empower doctors to explicitly project possible future trajectories and states of patients, allowing them to test the hypothetical effects and implications of different scenarios (e.g., realisation of selected parameters) as well as aid them in comprehension and comparison of these pathways. %
Notably, such a simulation-oriented system is capable of embracing %
\emph{missing} and \emph{unknown patient information} -- instead of ``solving'' this challenge via, potentially dubious, technical means -- by outputting alternative trajectories (and their likelihood) conditioned on different values of a selected variable, e.g., a medical test result. %

More broadly, this conceptualisation of artificial intelligence is able to support, and explicitly encourage, %
planning for multiple probable (future) outcomes rather than a single one chosen (possibly without robust justification) either based on an AI recommendation or a personal conviction. %
Recognising the contingency of different courses of action upon %
major critical junctures and branching points %
is beneficial as it promotes decisions that remain valid for the highest number of potential scenarios as well as solutions that can be easily adapted to unlikely situations~\citep{helmer1959epistemology,sokol2023navigating}. %
Such an AI tool %
could allow %
doctors to account for the uncertainty of the clinical environment as well as its dynamics, thus %
increase the robustness of their decision making by prompting them to explicitly consider the possible deviations, complications and unexpected events at each step of a patient pathway. %
This type of technological support %
has the potential to %
decrease the cognitive fatigue of clinicians, reducing their overconfidence and errors as well as improving their performance and decision consistency (thus overcoming bias and noise) in the demanding, high-paced and stressful healthcare setting. %

In addition to directly benefiting clinical practice, the simulation capability of AI tools can also be used for \emph{training} in a safe decision-making environment, offering plentiful chances for mental rehearsal, practice and learning, e.g., by replaying landmark case studies~\citep{kuhn2002diagnostic}. %
This functionality %
may alternatively be used to %
provide access to \emph{digital twins of individual clinicians} that can be consulted for ``second opinions'', compared across to identify points of disagreement leading to new insights, or used as benchmarks for junior doctors to learn from. %
This particular application of AI can prove especially potent when dealing with heterogeneous patient populations -- as is the case for paediatric sepsis -- where some clinicians may unknowingly exhibit better judgement for selected demographics. %
Combining many such independent ``opinions'' also facilitates %
\emph{the wisdom of the crowd} approach, which %
is a strategy for estimating an answer to a question by polling diverse individuals whose aggregate response is a better approximation of the ``correct'' answer. %
A specific instantiation of this process called the \emph{Delphi method} %
can be used by a group of people %
to arrive at a consensus; %
for example, clinicians rely on this technique to define and characterise complex medical conditions such as sepsis~\citep{goldstein2005international,singer2016third,schlapbach2018defining,morin2022current,schlapbach2024international}. %

\section*{Conclusion}

Our \emph{Perspective} proposes a radical shift in how artificial intelligence should be operationalised, away from the narrow focus on pursuing superhuman or state-of-the-art performance -- however defined -- and instead towards supporting people's cognitive and epistemic functions such as perception, reasoning and decision making. %
This view deprives AI of any special status or autonomy and intentionally decentres it, %
positioning this technology just as a supporting tool at the disposal of humans. %
Exactly thirty years after Collen concluded in his historical survey of medical informatics that %
``developing a comprehensive medical information system [appears to be] a more complex task than putting a man on the moon had been''~\citep{collen1995history} %
we are witnessing an explosion in AI's capabilities, %
yet on many planes its current conceptualisation does not seem to bring us any closer to this goal. %
The vision put forth by this \emph{Perspective} %
-- far from exhaustive in itself -- %
presents a different avenue for implementing artificial intelligence in medicine and beyond; %
one that unlocks capabilities and benefits %
unavailable when a data-driven model simply \emph{decides} and its explainability \emph{legitimises}. %

Built atop insights from cognitive sciences, our \emph{Perspective} seeks to seamlessly integrate AI tools %
into the broader systems ecology, %
aligning them %
with human needs and expectations as well as established real-life decision-making protocols and organisational workflows, taking care not to disrupt these (often fragile) structures. %
Instead of replacing people with undesired, fallible and potentially harmful automation, %
this novel approach %
empowers humans to make the best judgement given available information. %
In particular, it aims to %
promote best (clinical) decision-making practice and %
alleviate common reasoning shortcomings (e.g., arising due to cognitive biases), %
making these processes more factual, evidence-based and principled. %
It strives to achieve these goals %
by minimising decision errors as well as improving the consistency of and eliminating any undesired variability (i.e., noise) in human judgement. %

To this end, our \emph{Perspective} %
envisions AI assisting clinicians, and more generally the broader population, in everyday cognitive and epistemic tasks with which they commonly struggle or that exceed their capabilities. %
Providing doctors with %
timely and relevant (data-driven) insights %
can %
complement their expertise, boost their effectiveness %
and increase the likelihood of them reaching a correct diagnosis, %
leading to an overall improvement in decision making. %
To enhance and augment the abilities of clinicians, AI could, among others, %
illustrate the objective risk of various factors, present short- and long-term consequences of specific actions and communicate past (factual) decisions and their outcomes (e.g., to combat the time preference bias and undesired judgement variability). %
Artificial intelligence could also %
support clinicians in hypothesising about the effects and implications of different scenarios -- accounting for the uncertainty of the environment -- %
including any unlikely complications that they ought to anticipate, %
yielding more robust diagnoses. %

The main purpose of these algorithmic insights is to help doctors %
better understand what cues they need to look out for and what interventions they should consider implementing %
with the aim of alleviating the symptoms, improving the health state and managing the trajectories of their patients. %
Given the source and nature of this information, %
the task of interpreting and contextualising it remains strictly within the clinicians' remit.
Its safe incorporation into diagnostic reasoning processes is therefore the responsibility of individual doctors, as is currently the case with non-AI medical devices. %
Facilitating such integration of systems based on artificial intelligence, however, %
requires the underlying predictive models %
to be sound, reliable and robust %
so that they may be used as tools whose inner workings can remain opaque to their operators~\citep{sokol2023reasonable}; %
ante-hoc interpretability appears to offer a promising AI paradigm to achieve this goal~\citep{rudin2019stop,rudin2022interpretable}. %
More prosaic technological considerations include %
application-specific technical and operational requirements, e.g., availability of clinical variables for real-time deployment. %

Clearly,
advancing such systems along \emph{all} their sociotechnical dimensions is most likely to %
help them
overcome the pervasive translational barrier found in medical artificial intelligence research. %
Doing so within the framework introduced in this \emph{Perspective} is particularly promising. %
First and foremost, the key tenets of our approach agree %
with core beliefs and views of practising clinicians about the use of artificial intelligence in medicine~\citep{rober2024clinicians}. %
Our conceptualisation of AI is also crafted to %
closely align with the aforementioned ``social and relational rather than autonomous'' view of (human) intelligence. %
Additionally, our envisaged operationalisation of AI positions it as a partner for people to collaborate with or a tool at their disposal rather than their competitor or replacement, suggesting its favourable reception. %
Lastly, our approach keeps the allocation of decision responsibility with humans and preserves their autonomy, attesting to its strong sociotechnical foundations. %
All of these principles -- firmly grounded in decades of interdisciplinary research -- promise to increase acceptability and facilitate seamless integration of artificial intelligence in clinical practice for real-world impact.

When deployed and used, %
these systems can reduce mortality and morbidity of various diseases (besides alleviating their economic burden) by personalising treatments as well as minimising diagnostic and decision-making errors. %
Regarding paediatric sepsis, they could~\citep{tennant2024scoping}: %
\begin{itemize}
    \item 
aid with its \emph{detection} in view of this population's heterogeneity and the uncertainty surrounding this disease; %
    \item 
improve its \emph{management} given the lack of suitable (data-driven predictive) tools to assess its severity and monitor its progression; and %
    \item 
make its \emph{treatment} more consistent to reduce unnecessary exposure to antibiotics. %
\end{itemize}
To this end, such AI systems could, for example, be integrated into the fabric of multidisciplinary \emph{sepsis teams}~\citep{simon2022improved} and (bedside) \emph{sepsis huddles}~\citep{currie2023impact}, supporting the epistemic and cognitive functions of their participants. %
More broadly, these tools have the potential to generate new (scientific) knowledge and insights in (bio)medicine and healthcare as well as other disciplines amenable to AI modelling. %

\paragraph{Acknowledgements}
This research was supported by the Hasler Foundation, grant number 23082. %

\paragraph{Author Contributions}
\emph{Conceptualisation:} K.S. %
\emph{Methodology:} K.S. %
\emph{Formal Analysis:} K.S. %
\emph{Investigation:} K.S. %
\emph{Writing -- Original Draft:} K.S. %
\emph{Writing -- Review \& Editing:} K.S., J.F., J.E.V. %
\emph{Supervision:} J.F., J.E.V. %
\emph{Funding Acquisition:} K.S. %
All authors have read and approved the manuscript. %

\paragraph{Data Availability}
No datasets were generated or analysed in this study.

\paragraph{Competing Interests}
The authors declare no competing interests.

\bibliography{main}%

\end{document}